\documentclass[12pt]{article}

\newcommand{\be}{\begin{equation}}
\newcommand{\ee}{\end{equation}}

\begin{document}
\title{Information Storage in Black Holes}
  
\author{M. D. Maia\\
The Institute of Physics, University of Brasilia\\ 70919-970,
Bras\'{\i}lia, DF,  maia@unb.br}
\maketitle
\begin{abstract}
The information loss paradox for  Schwarzschild black holes   is examined,  using  the  ADS/CFT correspondence  extended to  the  $M_6 (4,2)$ bulk. It is found that  the only option compatible with the preservation of  the quantum unitarity is when   a  regular  remnant region  of the black hole survives to the black hole  evaporation process, where  information  can be  stored and eventually  retrieved.  
\end{abstract}

\begin{center}
{\large The  ADS/CFT  Correspondence}
\end{center}

 Conformal  covariance  in physics  was   firstly  introduced in electrodynamics back in  1909, but its  applications to fundamental physics 
 are  still  lagging \cite{Fulton}.
  Only very recently   the  ADS/CFT correspondence,   between   conformal fields in the Minkowski space  $M_4$    and gauge field theory  on the five-dimensional  deSitter gravity $AdS_5$,  started to make an impact on some relevant  issues related to  the  quantum  field  theory  on curved  space-times	\cite{Maldacena}.

  The  ADS/CFT  correspondence  can be  easily
   understood  when it is  seen  from   the point of  view   of   the  theory of  sub-manifolds.  Indeed, the four-dimensional anti-deSitter space-time  $AdS_4$ is   a hypersurface of negative constant  curvature embedded in the flat space $M_5 (3,2)$ \cite{Rosen}. By simply adding   one  extra   space-like   dimension, we  obtain    the five-dimensional negative constant curvature  surface   $AdS_5$ 
 as  a  hypersurface embedded    in the flat space $M_6 (4,2)$.    As it happens,  the 15-parameter conformal group   $C_o (3,1)$ acting on Minkowski's space-time  is  isomorphic  to the  group  $SO(4,2)$.   
  Since  $AdS_5$ is  a maximally symmetric  sub-space  of $M_6 (4,2)$,  it follows that any  (pseudo)  rotation  on  that  sub-space  corresponds  to  a conformal transformation  in $M_4$.   Thus,  for  a given   conformally invariant  field defined in Minkowski's space-time,  in  principle there  should be   a  corresponding   isometrically invariant field  in   the  $AdS_5$  space. 
    
  The  construction of the correspondence itself  may be not so  simple  as it depends on the nature and properties of the fields in  question.
  For example, we know how difficult it is to exactly solve  the non-Abelian Yang-Mills equations  in $M_4$. However,   in the  five-dimensional $AdS_5$  space   there exists  a particular   solution  for  the super-symmetric gauge-field, associated with the internal five-sphere $S^5$.  This solution appears in the derivation of the    heterotic  $E_8 \times  E_8$ string theory in the  10-dimensional  space $AdS_5 \times  S^5$ from  M-theory, through  a $Z_2$-symmetric orbifold compactification of the  $11^{th}$  dimension \cite{HW}. Therefore,  we obtain  a  super-conformal Yang-Mills field in  $M_4$ corresponding to  a  super-symmetric Yang-Mills field  in  $AdS_5$. Reciprocally,   the ADS/CFT  correspondence  can be used to explain how  the  unitarity   of  the quantum super-conformal Yang-Mills  field theory in  $M_4$, also  holds  in the five-dimensional $AdS_5$ gravitational environment \cite{Hawking2004}.
    
Four-dimensional gravity
can be recovered from this scheme as  a  brane-world of the 
 $AdS_5$ bulk. Accordingly, the  equations of  motion of the brane-world result from Einstein's  equations for the bulk geometry, together with   the four-dimensional  confinement  of  gauge fields,  and with the  exclusive probing of the extra dimensions by  gravitons. Although the full degrees of freedom required by the dynamical evolution of the brane-world may require more than five dimensions  \cite{ADD}, a  particular (albeit  very popular) model uses the  $AdS_5$  bulk by imposing  the same  $Z_2$   symmetry  across   fixed   boundary brane-worlds \cite{RS}. 
 
 However, the probing of the extra dimensions by gravitons should be  consistent   with  the classical  limit of the  future theory of quantum  branes,   generating 
   small deformations (or perturbations)  of the classical brane-world geometry. Coincidently, the theory of  sub-manifolds 
  developed  by   J. Nash describes such perturbations of   geometry,  requiring only that    the embedding functions  should  remain differentiable and regular \cite{Nash,Greene}. The regularity is  relevant to the application of the   inverse functions theorem, allowing us   to  keep a local  correspondence between  the original Riemannian  and the embedded sub-manifold. The problem appears  when  we try to apply Nash's  theorem  to   perturbations of a $Z_2$-symmetric brane-worlds  embedded in the   $AdS_5$ bulk. We find that  the  result is  not always compatible   with   the required regularity,
essentially   because each perturbation of a given manifold would have  a  mirror image  on the opposite side,
leading to an ambiguous  definition of its tangent vectors \cite{MaiaGDE}.  

 Another  difficulty facing a  five-dimensional  bulk with constant curvature
 is that  it  poses severe restrictions  to the  embedded brane-world.  For  example, in  the particular case of  a sub-manifold  with a
 spherically symmetric  metric, the  extrinsic  curvature turns  out to be  proportional to the metric: $k_{\mu\nu}= \alpha_0\; g_{\mu\nu}$, where  $\alpha_0$ denotes an integration  constant, characterizing umbilical points. With this, the  line element of   a  spherically  symmetric four-dimensional vacuum brane-world in a  five-dimensional constant curvature bulk  is like
 \begin{eqnarray*}
ds^2\!\! =\!\! (1\!\! -\! \frac{2m}{r}\!+\beta_0^2 r^2)^{-1}dr^2  +r^2 d\theta^2 
+ r^{2}\sin^{2}\theta d\phi^2
-(1\!\!-\!\!\frac{2m}{r} + \beta_0^2 r^{2})dt^2 
\end{eqnarray*}
where   $\beta_0^2 =( 3\alpha_{0}^2 -\Lambda_* )$  and where the    cosmological constant $\Lambda_*$  results from
 the  bulk's  constant curvature. Even considering 
$\Lambda_* =0$,  the constant $\alpha_0$  cannot  be  zero, under the penalty of  producing  just a trivial  solution $k_{\mu\nu}=0$ (which means a plane!).  The consequence is  that  the  good old  vacuum  Schwarzschild's  solution does not appear  in  the  $AdS_5$  bulk, but we may  have  the  Schwarzschild-Anti-deSitter solution.

\begin{center}
 { \large Extending the ADS/CFT  Correspondence}
\end{center}

 Since  the  $AdS_5$ and   $M_6 (4,2)$ have the same 15-parameter group of isometries,  all  arguments  based  on the  symmetries  of  the   $AdS_5$,
 including the  correspondence with conformal group $C_o (3,1)$, can be  extended to the whole space $M_6 (4,2)$. The  advantage of such extension is that $M_6 (4,2)$  contains  embedded sub-manifolds, including the Schwarzschild solution, which are not   present  in the  $AdS_5$  space.  Furthermore, at the points of  $M_6 (4,2)$ where  the embedding is regular, at least some of the  properties of  conformal fields in  $M_4$ can be  extended to these sub-manifolds,   and not just to the   $AdS_5$ space.
 
   As an example, twistors  are  usually defined  as conformally covariant spinors  on  $M_4$, but   they   can   also be  seen  as    spinors  defined by representations of the Clifford algebra in  $M_6 (4,2)$ \cite{Penrose,Murai}.
Therefore, we may extend some of the properties of twistors to the curved space  $AdS_5$, and  also to any   regular   vicinity of  the  Schwarzschild solution  embedded in $M_6 (4,2)$.
 
 This   extended ADS/CFT correspondence can also be used to   ensure that the   unitarity of the  quantum super-conformal  Yang-Mills theory in  $M_4$ can be extended to    points of  $M_6 (4,2)$ located
near the Schwarzschild's  black hole, by means of the regular embedding functions.
 Using such extension,  we may apply  the  same  arguments  of the 2004  version  of the 
information loss  theorem  to the standard  Schwarzschild black hole, as it appears in the 1975 theorem, instead of  depending only on the presence of extremal black holes in the 
$AdS_5$ bulk. Therefore, admitting that the   geometry  of the  brane-world   remains classical, we look again at the  three   possible  outcomes described at the conclusion of the original theorem  \cite{Hawking1975}: 
  
(a)  \emph{After the evaporation  everything disappears,  leaving  empty space  with no  singularity  at the origin}.  
 
 As in the    1975 paper, all information  is lost during the black hole thermal evaporation process, and  after  the evaporation anything that remains is  an  empty subspace of  the bulk.  On the other hand,   admitting that the unitarity is preserved under the  extended ADS/CFT  correspondence, some information should be retrievable. However,  the complete  evaporation  would imply  also in a   change from the multiply connected  to the simply connected topological configuration of  the sub-manifold geometry,  conflicting  with Geroch's theorem on the maintenance of  topology in a  classical geometry \cite{Geroch}.

 (b) \emph{ After the evaporation   a   naked  singularity  remains at $r=0$}. 

 The presence of the  singularity at  $r=0$, means  that after the evaporation the  Schwarzschild  space-time  has been  transformed into  its maximal  analytic  extension \cite{Fronsdal}, implying that 
   the   signature of  the bulk  change  from  $M_6 (4,2)$  to  $M_6 (5,1)$ \cite{Kruskal}. Consequently,   the  bulk symmetry  $SO(4,2)$ changes  to $SO(5,1)$,  and both   the standard and the extended ADS/CFT  correspondences  break  down. Therefore, the  unitarity of the conformal  quantum field near the  naked singularity can  no longer  guaranteed, at least  by use of such correspondences.

(c)\emph{ After the evaporation there is  a  remnant of the  black hole in a region  with a  finite diameter defined by  the fundamental  energy scale}.  

 In this scenario the topology  of the brane-world and the signature of the bulk are not affected by the evaporation, so that   the  extended ADS/CFR  correspondence and the  unitarity are preserved. Furthermore, admitting that  the  remnant   is  a regular  manifold  embedded in the  $M_6 (4,2)$  bulk,
 it may act as  an information  storage space. That is, the information that passes through  the  horizon and falls into the 
  remnant,  may eventually be retrieved by use of the  local   inverse  of the embedding map.
  This  regular remnant  is essentially  the same region  envisaged by Hawking in the  1975 paper, with the difference that  here it  is described as a regular sub-manifold embedded in   $M_6 (4,2)$, whose radius of curvature is  defined by the  fundamental energy scale.
\vspace{3mm}
  
  Although this extension of  the  ADS/CFT correspondence  may be  useful  to  understand the behavior of conformal  fields  near  curved  space-times,   it also implies  in   the emergence of   the 11-dimensional  space  $M_6 (4,2) \times  S^5$,  whose meaning, perhaps  in the context of  M-theory,  still remains to  be understood.

\end{document}